\documentclass[11pt]{article}

\usepackage[T1]{fontenc}
\usepackage[utf8]{inputenc}
\usepackage[margin=1in]{geometry}

\usepackage{amsmath,amssymb,bm}
\usepackage{graphicx}
\usepackage[numbers,sort&compress]{natbib}
\usepackage{microtype}
\usepackage{url}
\usepackage{xcolor}

\usepackage[
    colorlinks=true,
    citecolor=red,
    linkcolor=blue,
    urlcolor=blue,
    filecolor=blue
]{hyperref}

\usepackage{placeins}

\makeatletter

\renewenvironment{figure}[1][!t]
  {\@float{figure}[!t]}
  {\end@float}

\makeatother

\DeclareUnicodeCharacter{00D7}{\ensuremath{\times}}
\DeclareUnicodeCharacter{2248}{\ensuremath{\approx}}
\DeclareUnicodeCharacter{2265}{\ensuremath{\ge}}
\DeclareUnicodeCharacter{2212}{\ensuremath{-}}
\DeclareUnicodeCharacter{2208}{\ensuremath{\in}}
\DeclareUnicodeCharacter{2032}{\ensuremath{\prime}}
\DeclareUnicodeCharacter{0303}{}
\DeclareUnicodeCharacter{210E}{\ensuremath{h}}
\DeclareUnicodeCharacter{A700}{\ensuremath{_c}}
\DeclareUnicodeCharacter{208B}{\ensuremath{_-}}
\DeclareUnicodeCharacter{0304}{}
\DeclareUnicodeCharacter{0302}{}

\graphicspath{{figures/}}

\newcommand{\figureimage}[1]{%
    \includegraphics[
        width=\linewidth,
        height=0.75\textheight,
        keepaspectratio
    ]{#1}%
}

\newcommand{\paperfigure}[1]{%
    \IfFileExists{figures/fig_#1.pdf}{%
        \figureimage{figures/fig_#1.pdf}%
    }{%
        \IfFileExists{figures/fig_#1.png}{%
            \figureimage{figures/fig_#1.png}%
        }{%
            \IfFileExists{figures/fig_#1.jpg}{%
                \figureimage{figures/fig_#1.jpg}%
            }{%
                \fbox{%
                    \parbox[c][45mm][c]{0.9\linewidth}{%
                        \centering
                        Figure #1 placeholder\\
                        \small
                        Upload \texttt{figures/fig\_#1.pdf}
                    }%
                }%
            }%
        }%
    }%
}

\title{Inferring Experimental Objectives in Microscopy
Experiments from Decision Traces}

\author{
\normalsize
Kamyar Barakati$^{1,*}$,
Boris N. Slautin$^1$,
Aditya Raghavan$^1$,
Babita Babita$^2$,\\
\normalsize
Surya Prakash Reddy$^1$,
Jordan Marshall$^1$,
Munireach Nannory$^3$,
Stephen Puplampu$^3$,\\
\normalsize
Alla Slautina$^1$,
Faith M. Hines$^1$,
Newsha Javanmardi$^1$,
and Sergei V. Kalinin$^{1,2,\dagger}$\\[6pt]
\small
$^1$ Department of Materials Science and Engineering,
University of Tennessee, Knoxville, TN, USA\\
\small
$^2$ Department of Electrical Engineering and Computer Science,\\
\small
University of Tennessee, Knoxville, TN, USA\\
\small
$^3$ Department of Civil and Environmental Engineering,\\
\small
University of Tennessee, Knoxville, TN 37996, USA\\[4pt]
\small
$^*$ \texttt{kbarakat@vols.utk.edu}
\quad
$^\dagger$ \texttt{sergei2@utk.edu}
}

\date{}

\begin{document}

\maketitle

% ==================================================
% 9. Abstract
% ==================================================

\begin{abstract}
    Modern microscopes can acquire, analyze and act on data far faster than human operators can decide what to measure next. Exploiting this speed in automated experiments requires the instrument to follow the scientific logic of the operator, and one way to encode that logic is through reward functions that become part of the operation process. Here we explore whether such reward functions can be learned from the decision histories of microscope operators. We propose multiple frameworks in which recorded choices among alternative measurement locations serve as comparative evidence about the operator’s preferences, and we compare several algorithms for reconstructing the reward including interpretable linear models built on physically meaningful image descriptors, nonlinear functions of the same descriptors, an unsupervised image representation followed by preference modeling, and a preference-trained deep-kernel-inspired model. The result of this analysis is a spatial field of reward functions over the sample that can be used to run the instrument. We also propose a way to benchmark these approaches, based on withheld decisions and on decision histories generated with known objectives. We then examine the possibility that a decision history contains changing priorities throughout the experiment. A hidden Markov model assigns distinct rewards to recurring selection strategies, whereas a continuous-state model tracks smoothly varying preference weights. Controlled sequences generated over real microscopy candidate sets show that both concealed switches and gradual drift can be recovered. These reward functions, learned directly from recorded decisions, provide candidate objectives for future automated microscopy experiments.

\end{abstract}

% ==================================================
% 10. Main manuscript sections
% ==================================================

% Introduction
\section{Introduction}\label{sec:1}

Modern scientific instruments can acquire, analyze and act on data far faster than a human operator can make decisions. Scanning probe and electron microscopes can reposition, acquire images, and perform spectroscopy with varying acquisition times depending on the measurement technique, whereas a human operator may require minutes to decide where to measure next and why. Closing this gap is the key requirement for automated and autonomous experimentation, which has advanced rapidly through self-driving laboratories \cite{ref1,ref2,ref3,ref4,ref5,ref6}, machine-learning control of scanning probe microscopes \cite{ref7,ref8,ref9,ref10} and, most recently, agentic workflows that plan and execute multistep experiments \cite{ref11,ref12,ref13,ref14,ref15,ref16,ref17}.

An automated experiment, however, must know what it is looking for. Whether decisions are made by Bayesian optimization, reinforcement learning or (implicitly) by a language-model agent, they are ultimately driven by a reward function that assigns value to the possible measurements \cite{ref18,ref19,ref20,ref21}. For instrument optimization, the reward can be defined using an objective-specific measure of measurement quality. Bayesian optimization has been demonstrated in scanning probe microscopy to tune imaging parameters \cite{ref22,ref23} and in electron microscopy to optimize aberration-corrector settings \cite{ref24,ref25,ref26,ref27}. In our previous work, rewards were also formulated from physics and domain knowledge, favoring physically interpretable analyses, the discovery of structure–property relationships, and multi-objective combinations of image features \cite{ref28,ref29,ref30,ref31,ref32,ref33,ref34,ref35,ref36}. Such rewards are effective when the scientific objective can be written down in advance. Much of an experienced operator’s expertise, however, is tacit: experimentalists routinely discriminate among apparently similar regions based on observed contrast without formulating an explicit criterion.

Here we demonstrate how reward functions can instead be built from human decision-making itself. Every session at a microscope leaves a decision history in which, at each step, the operator chose one measurement among the alternatives available. In the theory of comparative judgment and discrete choice, such choices are evidence about an underlying utility \cite{ref37,ref38,ref39,ref40}, and inverse reinforcement learning and preference learning provide methods to infer that utility from behavior \cite{ref41,ref42,ref43,ref44,ref45,ref46,ref47,ref48,ref49,ref50,ref51}. If the alternatives are recorded together with the choice, the reward implied by the operator’s decisions becomes an inverse problem that can be solved from data, and the resulting function can be transferred to the instrument and executed at machine speed.

We first set up the problem of learning rewards from recorded alternatives and define how reconstructed rewards are evaluated (Section~\ref{sec:2}). Section~\ref{sec:3} reconstructs stationary rewards, progressing from linear models over interpretable image descriptors to nonlinear functions and learned image representations. Section~\ref{sec:4} considers decision histories in which the operator’s priorities change during the experiment and models them with discrete latent strategies and continuously evolving preferences. Section~\ref{sec:5} discusses what the reconstructed rewards reveal, their limitations, and how they can drive automated experiments.

% Learning rewards from recorded alternatives
\section{Learning rewards from recorded alternatives}\label{sec:2}

As an example, we consider ferroelectric domain structures in a \(\mathrm{PbTiO_3}\) thin film on a \(\mathrm{KTaO_3}\) substrate, a material whose ferroelastic a/c domain patterns, domain walls, and polarization switching have been studied extensively by piezoresponse force microscopy (PFM) \cite{ref54,ref55,ref56,ref57,ref58,ref59,ref60}. We analyze a single archived 256 × 256-pixel PFM image containing domain patterns, boundaries, and nonuniform textures. All recorded decision histories and controlled simulations are based on this image and its associated candidate patches. In a typical PFM workflow, the operator first acquires an image of the sample and then identifies locations for further examination, such as high-resolution imaging or spectroscopy.

Here, we focus on the retrospective interpretation of such an existing image and its recorded choice history. The record may represent selections made during instrument operation or evaluations subsequently performed over the archive; the inverse problem is the same provided the offered alternatives and the selected item are preserved. Importantly, the underlying preference need not remain stationary throughout the recorded sequence. A scientist may initially prioritize sharp domain walls, subsequently seek regions with ordered stripes, and later revisit a different morphology as the scientific question evolves. We ask whether one reward describes the complete record or whether changing preferences are identifiable from the same sequence of observed choices.

To study this problem, we consider the setting in which, at each step, the operator chooses one of 20 candidate segments of the image. We use this format because it matches what a human can do in practice: comparing a handful of alternatives immediately is fast and reliable, whereas ranking or scoring thousands of overlapping patches of a large image is not, and is easily biased toward the most visually salient features \cite{ref52}. At the same time, for a fixed reward function, the best of 20 randomly proposed candidates is close to the top of the reward distribution: on average, it lies above about 95\% of all patches of the image, allowing repeated selections to concentrate on high-value regions (Supplementary Section S2). Recording the alternatives together with the choice is essential because the winner and its 19 competitors tell us which features were preferred over which, information that a list of selected coordinates alone cannot provide. The resulting reward is an operational description of the operator’s preferences within this protocol rather than a physical law or a psychological model \cite{ref42,ref53}.

Using this image and selection protocol, we generated six decision histories, each consisting of 100 consecutive choices from sets of 20 candidate patches: two from human participants (P01 and P02), two from human-defined heuristics (HDH 1 and HDH 2), and two from a GPT-based language-model evaluator (GPT 1 and GPT 2). For the controlled tests in Section~\ref{sec:4}, new winner labels are additionally generated over the same candidate sets from simulated decision-makers with known image-dependent rewards.

Let the recorded alternatives at decision \(t\) be \(C_t=\{x_{t1},\ldots,x_{t20}\}\), and let \(a_{t}\) identify the chosen image patch. We fit a function \(R(x)\) that assigns each candidate a relative utility. A conditional SoftMax choice model maps these scores to selection probabilities \cite{ref38,ref39,ref40,ref61,ref62}; the inverse-temperature factor \(\beta \) describes how sharply the probability concentrates on higher-scoring alternatives. The likelihood is evaluated for the observed winner within its complete recorded candidate set. Because the observations constrain relative rather than absolute utility, rewards are identifiable only up to transformations allowed by the assumed choice model. Strongly correlated image descriptors and unobserved context create further ambiguity, discussed below and in Supplementary Section S2.

\begin{equation}
\label{eq:1}
P(a_t=i\mid C_t)=\frac{\exp[\beta R(\mathbf{x}_{t,i})]}{\sum_{j=1}^{20}\exp[\beta R(\mathbf{x}_{t,j})]}
\end{equation}

How well a reconstructed reward describes the operator is judged on decisions withheld from fitting. The simplest questions are whether the model ranks the actual winner first (top-1 accuracy) or among its first three choices (top-3 accuracy); a model choosing at random would score 5\% and 15\%. A finer measure is the probability the model assigns to the actual winner, summarized by the negative log-likelihood (NLL); random choice gives ln 20 ≈ 3.0, and a model that is confidently wrong is penalized more than a cautious one. Beyond these numbers, we can explore what the model has learned: which image characteristics it rewards, how stable the fitted coefficients are, how selected and rejected patches are arranged in a learned representation, and where the reward is high across the whole image. For models whose preferences change over time, we also ask how many distinct strategies the operator used, when the switches occurred, and whether each choice is predicted before the model is updated on it.

For \(T\) withheld decisions, the two principal metrics are

\begin{equation}
\label{eq:2}
\mathrm{NLL}=-\frac1T\sum_{t=1}^T\ln P(a_t\mid C_t)
\end{equation}

And

\begin{equation}
\label{eq:3}
\mathrm{Acc}_k=\frac1T\sum_{t=1}^T\mathbb{1}[r_t(a_t)\le k]
\end{equation}

where \(r_t(a_t)\) is the rank of the recorded winner among the 20 predicted scores and \(\mathbb{1}[\cdot]\) is the indicator function.

\begin{figure}[htbp]
  \centering
  \paperfigure{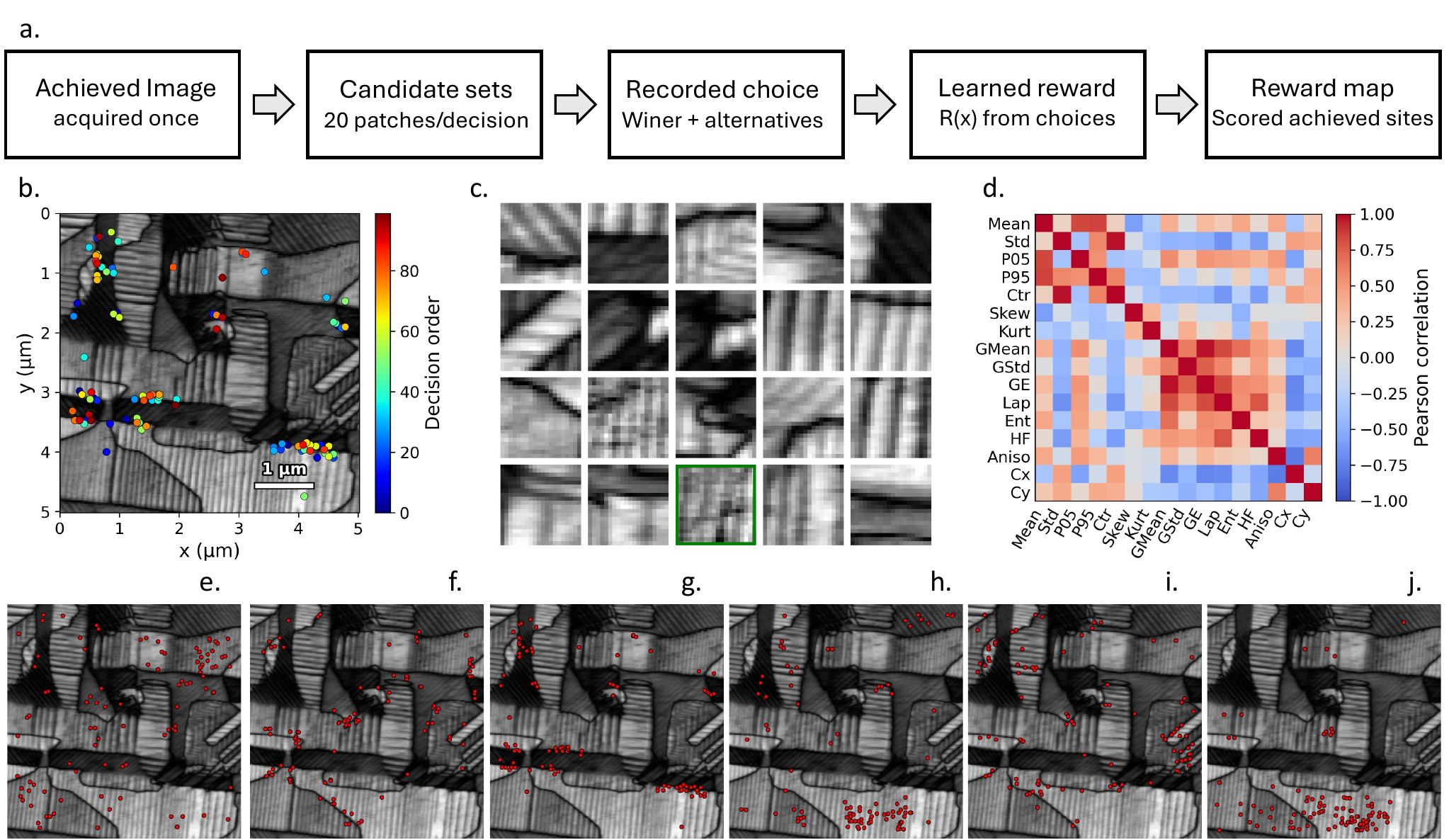}
  \caption{Retrospective preference reconstruction from microscopy decision traces. (a) Workflow: an archived image supplies limited candidate sets, the recorded choice and its alternatives constrain an inferred reward, and the frozen reward is projected over the same image. (b) Spatial distribution of recorded selections, color-coded by decision order. (c) One recorded set of 20 candidate patches; the winner is outlined. (d) Pearson correlations among the 16 image descriptors across the experimental decision traces. (e–j) Spatial distributions of 100 recorded selections for (e) GPT 1, (f) GPT 2, (g) HDH 1, (h) HDH 2, (i) P01, and (j) P02. Red markers indicate selected locations on the original PFM image.}
  \label{fig:1}
\end{figure}

The information supplied to the inverse problem is summarized in Figure~\ref{fig:1}. An archived image provides candidate sets, the recorded winner and its alternatives constrain a reward, and the frozen reward can be projected back over the whole image (Fig.~\ref{fig:1}a). Individual histories concentrate their selections on a few domain regions (Fig.~\ref{fig:1}b). Each decision offers 20 patches, of which one is selected (Fig.~\ref{fig:1}c). Several of the 16 image descriptors are strongly correlated, notably the intensity statistics and the gradient–Laplacian group (Fig.~\ref{fig:1}d). Different histories over the same image produce clearly different spatial patterns, which motivates learning a reward for each history rather than assuming one universal image criterion (Fig.~\ref{fig:1}e-j).

A reconstructed reward can be used at three levels. First, it should predict the choices that the operator made but that were withheld from fitting. Second, the frozen reward can score every other patch of the image, or of a new image, producing a reward map that can be inspected. Third, it can be used to choose new measurements on the instrument. In this paper we test the first level quantitatively, illustrate the second, and lay the groundwork for the third.

We compare several ways of reconstructing the reward rather than assuming that more sophisticated models are necessarily better. Two choices define each approach: how a patch is represented, either by physically interpretable descriptors or by features learned from the images, and which function converts this representation into a reward, from a weighted sum to nonlinear and kernel functions. We then ask whether the reward itself changes during a decision history, describing recurring strategies with a finite set of states and gradual changes with continuously updated weights.

% Stationary preference reconstruction
\section{Reconstructing rewards under a stationary-preference assumption}\label{sec:3}

This section reconstructs a single, time-independent reward from each decision history. We begin with linear rewards over physically interpretable image descriptors (Section~\ref{sec:3.1}), which make the inferred preference directly readable in terms of familiar image characteristics. We then keep the same descriptors but allow nonlinear additive and kernel functions (Section~\ref{sec:3.2}) and finally replace the handcrafted descriptors with learned image representations, first from an unsupervised variational autoencoder and then from a convolutional model trained on the choices themselves (Section~\ref{sec:3.3}). Each model is evaluated on withheld decisions with the metrics of Eq.~\eqref{eq:2} and \eqref{eq:3} and visualized as a spatial reward field over the archived image.

\subsection{Interpretable image descriptors: what characteristics of a patch explain selection?}\label{sec:3.1}

A transparent starting point is to express the utility of a patch through familiar image characteristics. Local contrast and intensity variation, gradient and Laplacian responses, boundary content, anisotropy, and spatial-frequency statistics all provide measurable descriptions of ferroic domain morphology. The first physics-guided formulation introduced six descriptors including contrast, boundary strength, orientation, periodicity, characteristic scale, and localization, and optimized a parametric score using a ranking margin \cite{ref63}. In this analysis, we use 16 image statistics and also test a reduced set of six. The reduced model is not identical to the earlier physics-guided six-feature reward: the descriptors, fitting objectives, and data splits are different. We therefore compare their behavior, rather than treating them as the same model.

Each candidate patch is represented as a bag of image features: a fixed-length vector of descriptors computed from the pixel intensities without reference to the position of the patch in the image \cite{ref64,ref65}. The descriptors are standardized with the mean \(\mu _{m}\) and standard deviation \(s_{m}\) of the training candidates,

\begin{equation}
\label{eq:4}
\boldsymbol\phi(x)=[\phi_1(x),\ldots,\phi_{16}(x)]^T,\quad \tilde\phi_m(x)=\frac{\phi_m(x)-\mu_m}{s_m}
\end{equation}

For a patch with intensities \(I\)(\(p\)) at pixels \(p\) ∈ \(x\), representative descriptors are the mean gradient magnitude and the mean absolute Laplacian,

\begin{equation}
\label{eq:5}
\phi_{\rm grad}=\frac1{|x|}\sum_{p\in x}\|\nabla I(p)\|,\quad \phi_{\rm lap}=\frac1{|x|}\sum_{p\in x}|\nabla^2 I(p)|
\end{equation}

the histogram entropy, with \(h_b\) the fraction of pixels in intensity bin b, and the fraction of Fourier power above a cutoff spatial frequency \(k_c\),

\begin{equation}
\label{eq:6}
\phi_H=-\sum_b h_b\ln h_b,\quad \phi_{\rm HF}=\frac{\sum_{|k|>k_c}|\hat I(k)|^2}{\sum_k|\hat I(k)|^2}
\end{equation}

and the anisotropy of the structure tensor \(J\) with eigenvalues \(\lambda_1\ge\lambda_2\) \cite{ref66},

\begin{equation}
\label{eq:7}
\phi_{\rm aniso}=\frac{\lambda_1-\lambda_2}{\lambda_1+\lambda_2},\quad J=\sum_{p\in x}\nabla I(p)\nabla I(p)^T
\end{equation}

which is zero for isotropic texture and approaches one for stripe domains of a single orientation. The remaining descriptors include intensity statistics, gradient statistics, and horizontal and vertical nearest-neighbor correlations. Their definitions and implementation details are summarized in Supplementary Section S3.

For standardized descriptors, a linear reward is the weighted sum given in Eq.~\eqref{eq:8}. This linear approach allows immediate interpretation \cite{ref67,ref68,ref69}: each coefficient states how strongly a given image characteristic attracts or repels the decision-maker. A positive coefficient means that patches with larger values of the descriptor are more likely to be chosen, and a negative coefficient means the opposite. Because several descriptors measure related properties, for example gradients and the Laplacian response, the weight can be shared between them in different ways that predict the same choices equally well. Individual coefficients should therefore be read together with the selected patches themselves and with the competitors they were preferred over.

\begin{equation}
\label{eq:8}
R(x)=\mathbf w^T\tilde{\boldsymbol\phi}(x)=\sum_{m=1}^{16}w_m\tilde\phi_m(x)
\end{equation}

The coefficients are estimated from the training decisions by minimizing the regularized choice likelihood,

\begin{equation}
\label{eq:9}
\hat{\mathbf w}=\arg\min_{\mathbf w}\left[-\sum_{t\in\mathrm{train}}\ln P(a_t\mid C_t;\mathbf w)+\frac\lambda2\|\mathbf w\|^2\right]
\end{equation}

where \(P\) is given by Eq.~\eqref{eq:1} and \(\lambda \) sets the strength of the \(L\)2 penalty. A positive coefficient indicates that, after controlling for the other descriptors, larger values accompany selected patches.

\begin{figure}[htbp]
  \centering
  \paperfigure{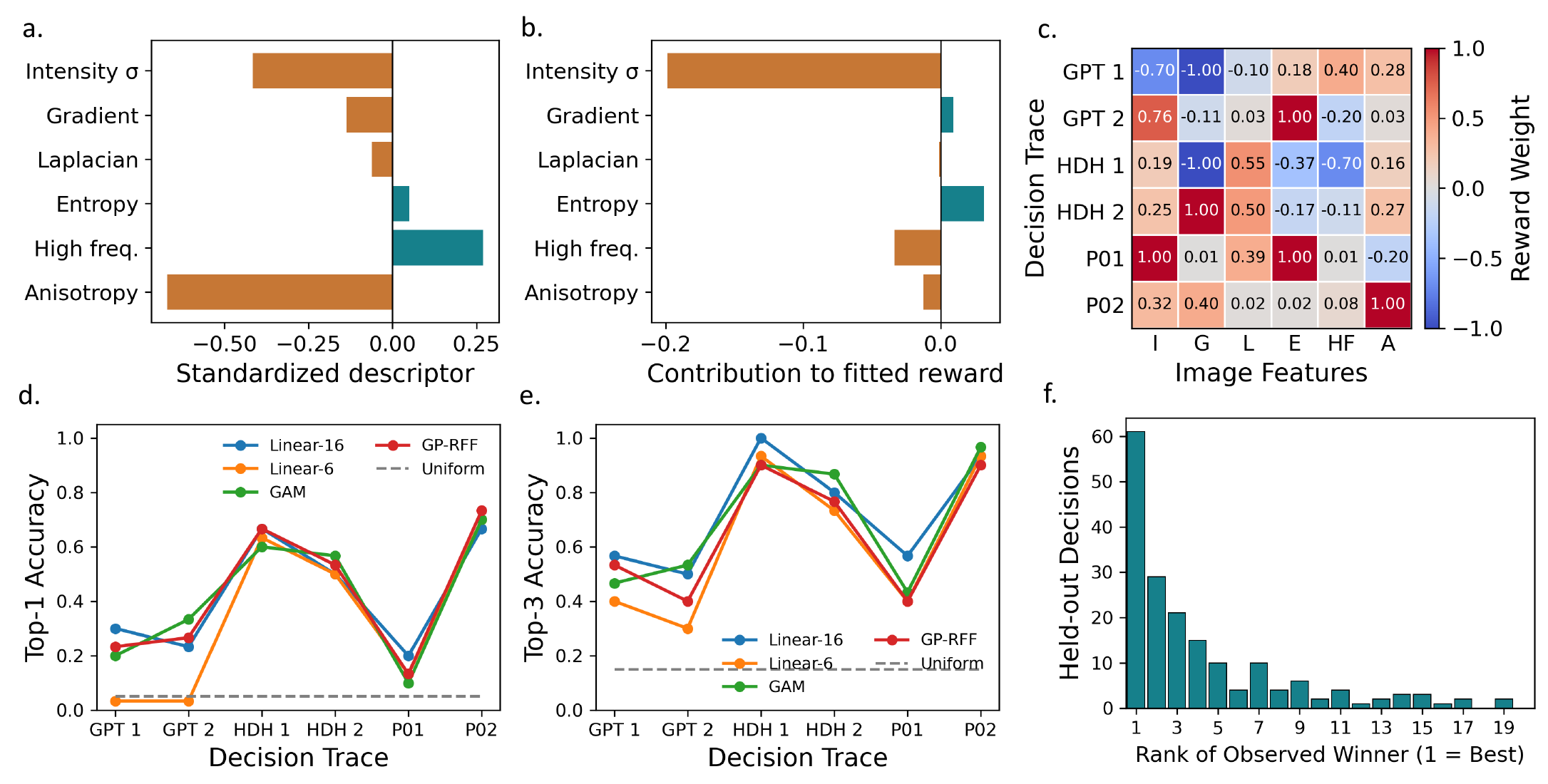}
  \caption{Interpretation and testing of descriptor-based rewards. (a) Standardized values of six image descriptors for a representative patch. (b) Their contributions (descriptor × fitted coefficient) to the linear utility. (c) Standardized coefficients of the 16-descriptor linear model for the six recorded decision histories: GPT 1, GPT 2, HDH 1, HDH 2, P01, and P02. (d, e) Held-out top-1 (d) and top-3 (e) accuracy for each trace and model; dashed lines mark uniform choice (5\% and 15\%). (f) Rank of the recorded winner among its 20 alternatives for the held-out decisions of a representative trace.}
  \label{fig:2}
\end{figure}

Figure~\ref{fig:2} follows the descriptor model from image measurements to predicted choices. For a representative patch, standardized descriptor values (Fig.~\ref{fig:2}a) combine with the fitted coefficients into contributions to a single utility function (Fig.~\ref{fig:2}b), here dominated by intensity spread. The coefficient matrix of the 16-descriptor model differs markedly between traces (Fig.~\ref{fig:2}c): the heuristic-driven histories show relatively strong contributions from selected image descriptors, whereas the GPT-generated histories exhibit more distributed coefficients. Held-out top-1 and top-3 accuracies range from chance to 90\% and 100\%, respectively (Fig.~\ref{fig:2}d, e), and for a well-reconstructed trace almost every held-out winner is ranked first or second among its 20 alternatives (Fig.~\ref{fig:2}f). Two parameter vectors can differ substantially yet rank the observed alternatives similarly, so these diagnostics are read together rather than from the coefficient map alone.

The resulting recoverability is highly trace dependent. The heuristic selection rules are reconstructed most accurately from image descriptors, consistent with those generating rules depending on visible patch statistics. Other histories, especially some GPT-generated choices, remain difficult for the same features. A low score has several possible explanations: the representation omitted a relevant image characteristic; choices were probabilistic or affected by unrecorded context; the underlying evaluator changed its objective; or the limited training set did not constrain the model sufficiently. The data do not distinguish these possibilities automatically. Likewise, high within-image accuracy does not establish that a human scientific objective has been uniquely recovered. It establishes predictive agreement with a particular recorded choice process under its candidate-generation protocol.

The metric chosen for monitoring changes the interpretation. A model may rank the actual winner second on many decisions and still capture a meaningful preference, whereas another may rank most winners first but assign an implausibly high probability to the wrong patch on its remaining failures. The distribution of selected-minus-best-rejected margins and the conditional rank of each winner therefore complement top-1 accuracy. We also check model convergence, how the fitted coefficients vary among traces, and how sensitive the results are to the held-out decisions. These checks are particularly important for more flexible functions, because an apparently smooth reward field can be produced by a model that poorly predicts the recorded consideration sets.

\subsection{Alternative functions over the same image descriptors}\label{sec:3.2}

There is no physical reason for experimental preference to be linear in an image statistic. Intermediate contrast may be preferable to either very low contrast or saturated contrast; a particular domain periodicity may be informative only within a narrow range; and the interpretation of a gradient depends on whether it belongs to an ordered boundary or noise. We therefore compare several functions defined over the same sixteen image descriptors while keeping the candidate sets and the inverse-choice objective fixed. This distinguishes two sources of modeling flexibility: the description of an image and the mathematical dependence of preference on that description. It also prevents an apparent advantage of a learned image encoder from being attributed automatically to a nonlinear reward head.

A generalized additive model (GAM) replaces individual linear terms with smooth univariate functions and can represent nonmonotonic dependence while retaining feature-level interpretability. A finite random-Fourier-feature approximation to a kernel reward permits interactions among descriptors but is not an exact Gaussian-process preference posterior. An iterative latent-utility method alternates between estimating utilities associated with individual candidate decisions and updating a systematic reward predictor; it is a regularized posterior-mode construction rather than exact expectation–maximization. These are alternative hypotheses about the form of an effective preference, not a hierarchy expected to improve monotonically. None predicts unmeasured polarization or spectroscopic response merely because it assigns higher utility to a patch, and their numeric scales cannot be compared across fits without a common identification convention. In the notation of Eq.~\eqref{eq:8}, the additive and kernel alternatives replace the linear score by

\begin{equation}
\label{eq:10}
R_{\rm GAM}(x)=\sum_m f_m(\tilde\phi_m(x)),\quad f_m(u)=\sum_b c_{mb}B_b(u)
\end{equation}

\begin{equation}
\label{eq:11}
R_{\rm RFF}(x)=\boldsymbol\beta^T\mathbf z(\tilde{\boldsymbol\phi}(x)),\quad \mathbf z(u)=\sqrt{2/D}\cos(\Omega u+\mathbf b)
\end{equation}

where \(B_{b}\) are spline basis functions with coefficients \(c_{mb}\), the \(D\) rows of \(\Omega \) are drawn from a Gaussian with inverse length scale \(l^{-1}\), and the phases b are uniform on [0, 2\(\pi \)), so that \(z(u)^Tz(u\prime)\) approximates a radial-basis-function kernel \cite{ref70,ref71,ref72}. Both models are fitted with the objective of Eq.~\eqref{eq:9}.

\begin{figure}[htbp]
  \centering
  \paperfigure{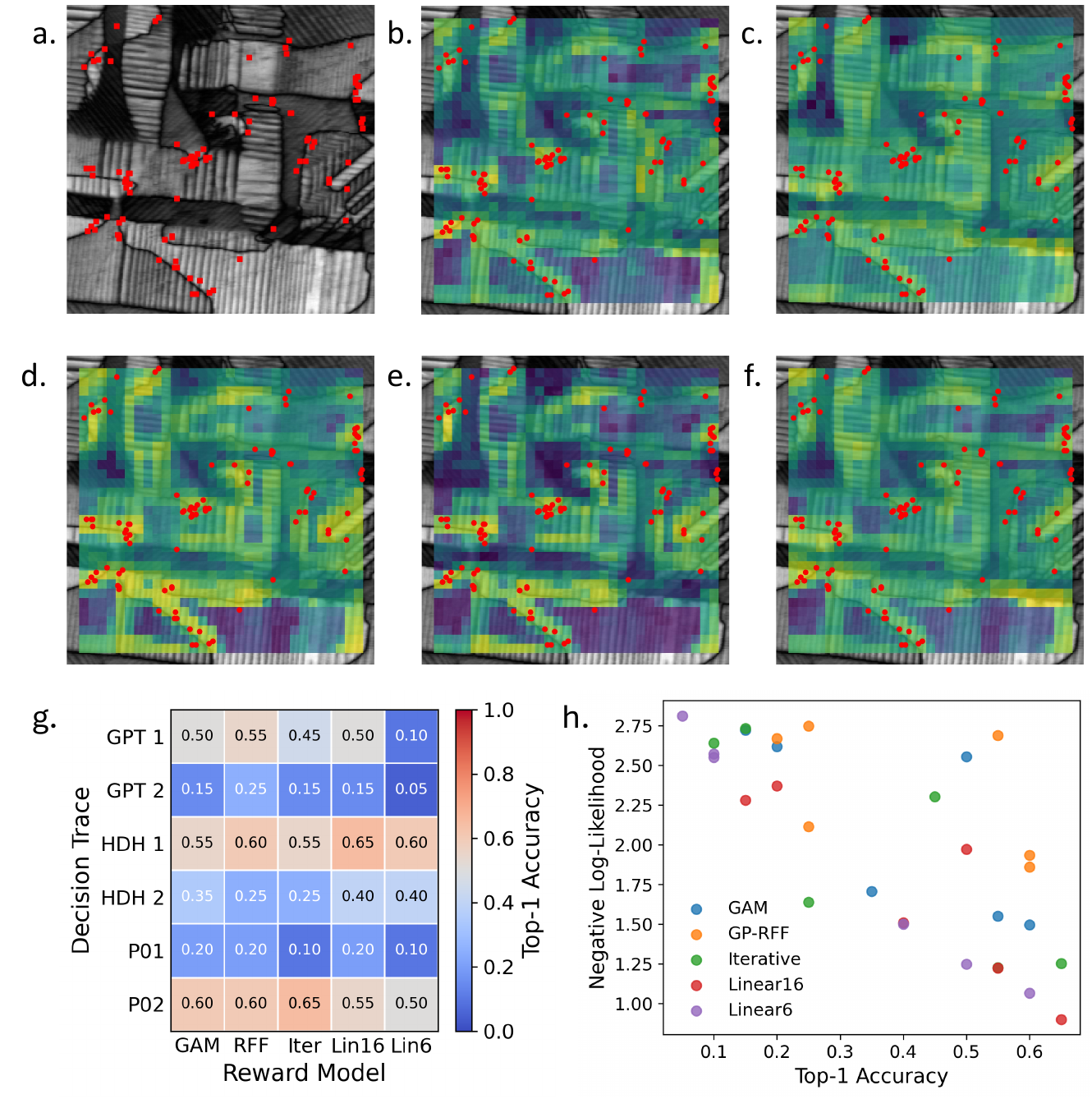}
  \caption{Spatial reconstruction and predictive performance of stationary reward models. (a) Recorded selection locations for GPT 1 overlaid on the original PFM image (red markers). (b–f) Spatial reward maps reconstructed from the same archived image using (b) the 16-descriptor linear model (Linear16), (c) the six-descriptor linear model (Linear6), (d) the generalized additive model (GAM), (e) the random Fourier feature kernel model (GP-RFF), and (f) the iterative reward model. Each reward map is displayed using an independently normalized color scale, with warmer colors indicating higher inferred rewards. Red markers indicate recorded selections, and the reward fields are evaluated over the valid patch-sampling region. (g) Held-out Top-1 accuracy of the five reward models across six decision traces: GPT 1, GPT 2, HDH 1, HDH 2, P01, and P02. (h) Relationship between held-out Top-1 accuracy and negative log-likelihood for all 30 trace–model combinations, illustrating differences between ranking accuracy and probabilistic prediction performance.}
  \label{fig:3}
\end{figure}

The six-trace benchmark fits each model independently and evaluates it on the original seeded 80/20 decision-level holdout. Figure~\ref{fig:3} juxtaposes the recorded selections of GPT 1 with the reward fields of all five models (Fig.~\ref{fig:3}a–f) and reports held-out top-1 accuracy for every trace and model (Fig.~\ref{fig:3}g), together with its relation to choice likelihood (Fig.~\ref{fig:3}h). The image-wide maps are derived by evaluating the frozen reward on overlapping patches from an already acquired parent image; they do not represent additional measurements. Predictions within the twenty alternatives available at each decision provide the primary test of retrospective learning. The spatial field instead asks what structures the fitted model would consider valuable if those patches were offered, and it may differ appreciably even among models with similar held-out accuracy.

Across the six recorded decision histories, the sixteen-descriptor linear reward model achieved a mean held-out Top-1 accuracy of 40.8\%, comparable to GP-RFF (40.8\%) and GAM (39.2\%), while the iterative and six-descriptor linear models achieved 35.8\% and 29.2\%, respectively (Fig.~\ref{fig:3}g; Supplementary Table S3). Performance varied substantially across traces, reflecting differences in the recorded selection strategies. These averages should not be interpreted as a definitive ranking of model architectures: each trace contains only 20 held-out decisions, so a single changed prediction shifts its accuracy by five percentage points, and all traces draw candidate patches from the same archived image. Nevertheless, the results indicate that relatively simple, physically interpretable image descriptors can recover a substantial fraction of observed selection preferences, while more flexible nonlinear reward functions do not consistently improve predictive performance.

Top-1 accuracy and likelihood can favor different models (Fig.~\ref{fig:3}h). Top-1 only asks whether the highest-scoring candidate is the actual winner, whereas the likelihood also credits a model for assigning reasonable probability to winners it does not rank first and penalizes it for confident mistakes. Training curves, the score distributions of selected and rejected patches, and the ranks of the actual winners help determine whether a poor result reflects missing image information, poor calibration or overfitting. Reward maps often highlight similar domain features even when the likelihoods differ, so the maps are best used to interpret a model, and the withheld choices to test it. This analysis yields a dense spatial reward map: every patch of the image receives a score, and the map can be interpreted directly in terms of the features the decision-maker preferred, such as sharp boundaries, periodic stripes or contrast variations. The same map can be used as an effective reward function for selecting new locations to sample. However, it is important to recognize its limitations. Neighboring patches overlap strongly, so a contiguous high-reward region is not a set of independent informative locations, and the operator never saw most of the high-scoring locations when making a given choice. The map therefore complements, but does not replace, the test on withheld choices. A frozen reward can also re-rank all other patches of the archived image, illustrating how an automated system would prioritize them. This is a projection of the learned objective rather than a record of an experiment that was performed; how such a reward is converted into a sampling policy is discussed in Section~\ref{sec:5} and Supplementary Section S6.

\subsection{Learning image representations instead of specifying all descriptors}\label{sec:3.3}

Handcrafted descriptors are transparent, but microscopy images contain junctions, domains of different sizes and long-range textures that a short list of statistics may not capture. For example, a human operator may prefer regions with curved domain walls rather than straight ones, even when their image statistics are nearly equivalent. An alternative is to let a neural network learn the image representation. This raises a simple question: is a representation learned to reconstruct the images good enough for predicting choices, or must the network be trained on the choices themselves? We test both options below, the first with a variational autoencoder and the second with a preference-trained convolutional model.

\begin{figure}[htbp]
  \centering
  \paperfigure{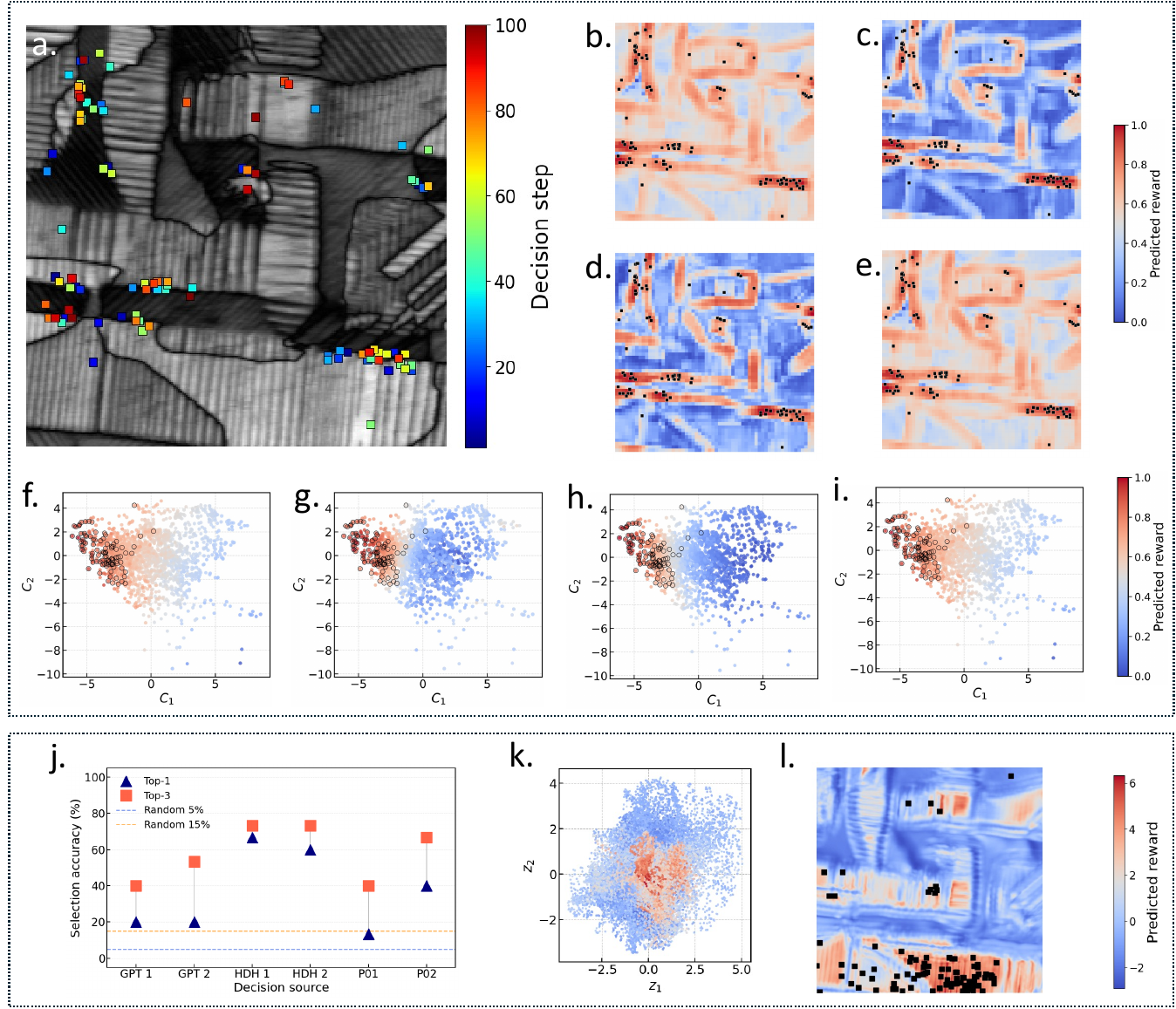}
  \caption{Reward reconstruction using a preference-trained deep-kernel-inspired model and variational autoencoder–Gaussian process (VAE–GP) representations. Upper block (a–i): Deep-kernel-inspired reconstruction of decision preferences. (a) Original PFM image with recorded selection locations from HDH 1, color-coded by decision order. (b–e) Spatial reward maps reconstructed for four representative decision histories, with black markers indicating recorded selections. (f–i) Corresponding reward landscapes projected onto the two-dimensional learned feature space, where the color scale represents the predicted reward and outlined markers indicate recorded selections. Panels (e, i) correspond to P02. Reward values are normalized independently for each model. Lower block (j–l): VAE–GP reward reconstruction. (j) Held-out Top-1 and Top-3 selection accuracies across the six decision traces (GPT 1, GPT 2, HDH 1, HDH 2, P01, and P02), with dashed lines indicating random-choice baselines of 5\% and 15\%, respectively. (k) Predicted reward landscape for P02 visualized in two dimensions of the VAE latent representation. (l) Corresponding spatial reward map projected onto the original PFM image, with black squares marking recorded selection locations.}
  \label{fig:4}
\end{figure}

In a variational autoencoder (VAE) \cite{ref73,ref104,ref105,ref106}, the encoder is learned from an image-reconstruction objective and latent regularization. The selected-candidate labels do not supervise this representation. Preference inference is subsequently performed in the latent space. For the six-trace collection, an eight-dimensional VAE was combined with a Gaussian-process preference model (Fig.~\ref{fig:4}j-l). A separate executable exploratory run used VAE coordinates with a finite random-feature utility approximation, which should not be described as a reproduction of the full GP posterior. This distinction matters scientifically: low reconstruction error does not guarantee that a latent direction captures domain structures that an operator considers interesting, and good reward prediction does not require faithful reconstruction of every pixel. The two objectives measure different aspects of the information retained from an image.

The encoder and decoder are trained by maximizing the evidence lower bound \cite{ref73},

\begin{equation}
\label{eq:12}
L_{\rm VAE}=\mathbb E_{q(z\mid x)}[\ln p(x\mid z)]-\mathrm{KL}[q(z\mid x)\|p(z)]
\end{equation}

and the encoder means then serve as inputs to the choice model of Eq.~\eqref{eq:1}.

A preference-trained convolutional model optimizes the image representation and the reward function jointly against the recorded winner choices. In the implemented deep-kernel-inspired construction, a convolutional encoder maps 20 × 20-pixel patches to two latent coordinates and a finite random-Fourier-feature head assigns utilities. The choice likelihood trains the representation to discriminate selected patches from their actual competitors. Although the architecture resembles deep kernel learning \cite{ref74,ref107,ref108,ref109,ref110}, the implemented finite-feature model should not be equated with a fully Bayesian Gaussian-process posterior: any final-layer Hessian or Laplace approximation conditions on the fitted encoder and does not propagate all image-encoder uncertainty. We therefore examine internal activations, latent geometry, reward distributions, and held-out choices rather than invoking a Bayesian uncertainty claim unsupported by the training procedure. Architecture and preprocessing details are in Supplementary Section S4.

The preference-trained reward composes a convolutional encoder g with parameters \(\theta \) and the random-feature head of Eq.~\eqref{eq:11} \cite{ref74},

\begin{equation}
\label{eq:13}
R_{\rm DKL}(x)=\boldsymbol\beta^T\mathbf z(g_\theta(x)),\quad g_\theta:\mathbb R^{20\times20}\to\mathbb R^2
\end{equation}

and \(\theta \) and \(\beta \) are fitted jointly by minimizing the choice NLL of Eq.~\eqref{eq:2} on the training decisions; the Laplace approximation \cite{ref75,ref76} is applied to \(\beta \) only.

Figure~\ref{fig:4} extends stationary reward reconstruction to learned image representations, comparing deep kernel learning (DKL) and variational autoencoder–Gaussian process (VAE–GP) approaches across the six decision traces. The upper panels (Fig.~\ref{fig:4}a–i) illustrate preference reconstruction using the deep-kernel-inspired model. Figure~\ref{fig:4}a shows the recorded selections for HDH 1, while Figs. (4b–e) and (4f–i) show the spatial and latent reward landscapes obtained by fitting the same model architecture independently to four representative decision histories. The recorded choices are concentrated near prominent domain boundaries and regions of strong local intensity variation (Fig.~\ref{fig:4}a). The reconstructed spatial reward maps (Fig.~\ref{fig:4}b–e) reveal distinct patterns of preference across the archived PFM image, with elevated rewards in several regions containing recorded selections. Differences in the spatial extent and intensity of these regions indicate that the independently fitted models capture different preferences across the recorded decision histories. Projection into the two-dimensional learned feature space further reveals how the models distribute rewards across the candidate-image representation (Fig.~\ref{fig:4}f–i). Recorded selections occupy a relatively restricted portion of this space, and regions of elevated predicted reward partially overlap with their distribution. These projections provide an interpretable visualization of the learned preference landscapes. For the preference-trained model, the two coordinates constitute the learned representation used for reward prediction, whereas the VAE–GP results display only two coordinates of the eight-dimensional VAE representation.

The lower panels (Fig.~\ref{fig:4}j–l) investigate the VAE–GP approach, in which a Gaussian-process preference model \cite{ref50,ref51} operates on image representations learned by a variational autoencoder. Across the six decision histories, the model places the recorded winner among its three highest-ranked candidates in approximately 73\% of held-out decisions for both heuristic-driven traces, 40\% and 67\% for participants P01 and P02, and 40\% and 53\% for GPT 1 and GPT 2, respectively (Fig.~\ref{fig:4}j). These results demonstrate that learned image representations can recover meaningful selection preferences, although predictive performance varies substantially across decision sources. For participant P02, the reconstructed reward landscape varies smoothly across the two-dimensional VAE latent space (Fig.~\ref{fig:4}k). Projection of this reward onto the archived PFM image reveals elevated preferences near regions of pronounced domain contrast, including areas containing clusters of recorded selections (Fig.~\ref{fig:4}l). However, because the VAE representation is learned through image reconstruction rather than candidate-selection supervision, visually prominent features may not necessarily correspond to the characteristics driving recorded decisions. In contrast, the preference-trained DKL approach directly optimizes the image representation using observed selections, potentially emphasizing features more relevant to the underlying decision strategy.

For HDH 1, the preference-trained model correctly ranks 14 of 20 held-out winners first (70\%), compared with 13 of 20 (65\%) for the corresponding linear descriptor model. However, the linear model achieves a lower NLL (0.90 versus 1.19). Thus, the preference-trained representation provides a modest improvement in top-1 accuracy for this trace without improving probabilistic prediction. The nonlinear representation therefore improves one count while worsening the probability assigned to the observed choices. This makes the example informative rather than universally favorable to deep models. A comparable supervision budget, feature preprocessing, repeated test partitions, and independent-image evaluation would be necessary to establish general superiority of one representation. In the archived benchmarks, some of the six traces remain difficult regardless of representation, suggesting that omitted context and nonstationary cannot be solved solely by expanding the network.

\begin{figure}[htbp]
  \centering
  \paperfigure{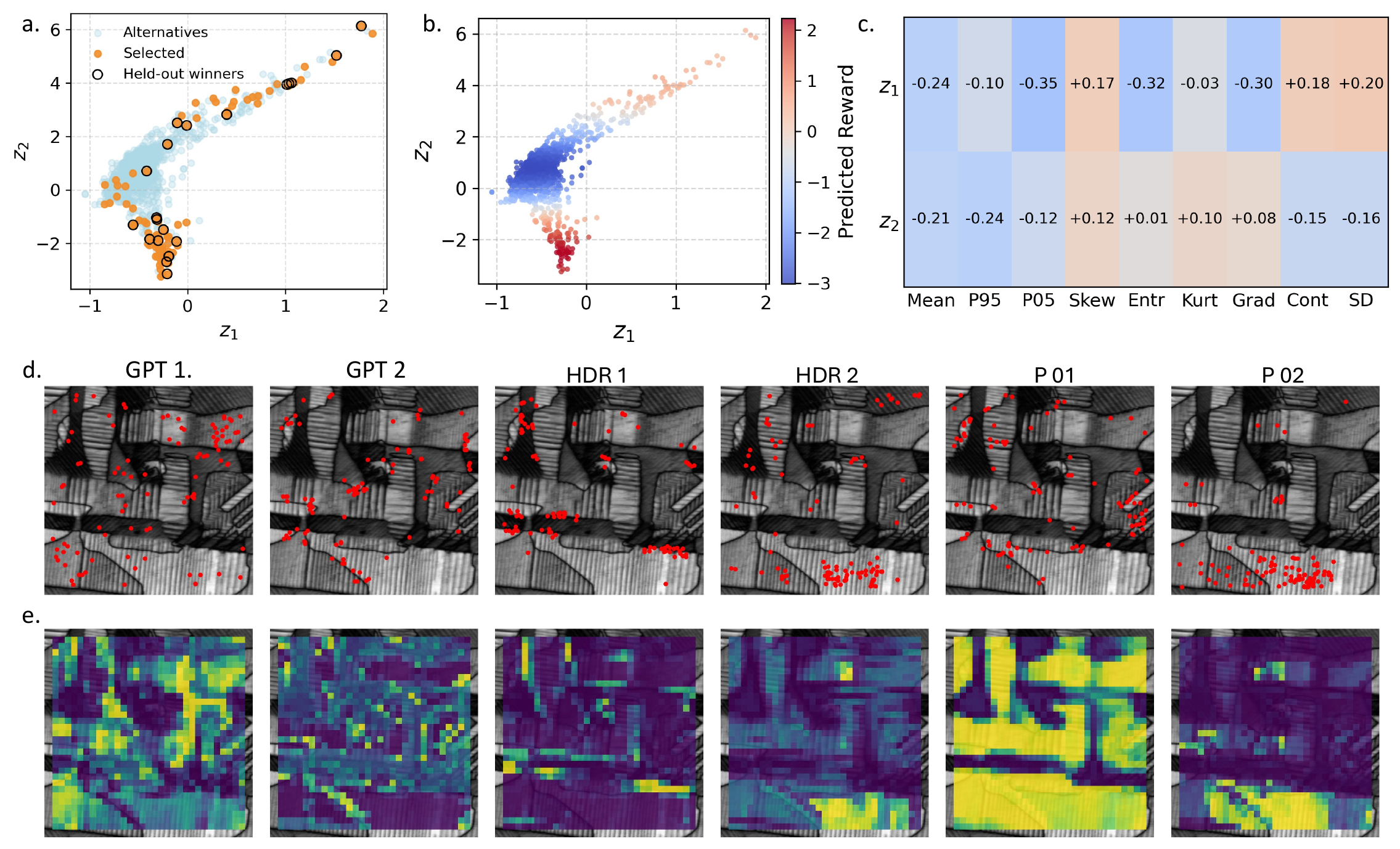}
  \caption{Preference-trained deep kernel learning (DKL) for reward reconstruction. (a) Learned two-dimensional embedding for HDH 1, showing selected candidates (orange), alternatives (light blue), and held-out winners (black circles). (b) Predicted reward across the same latent space. (c) Spearman correlations between latent coordinates and nine image descriptors. (d) Recorded selection locations for GPT 1, GPT 2, HDH 1, HDH 2, P01, and P02. (e) Corresponding spatial reward maps from independently trained DKL models, each displayed on its own color scale.}
  \label{fig:5}
\end{figure}

The two latent coordinates also admit a more physical test than a visually attractive scatter plot. Correlation against simple image statistics reveals whether the learned axes largely encode intensity, boundaries, periodicity, or other measured properties. In the illustration (Fig.~\ref{fig:5}c), intensity-related descriptors explain a substantial component of the latent variation, suggesting that even a deep preference model may rediscover relatively simple image regularities. These correlations help us understand what the latent coordinates represent, but they do not tell us why the decision-maker preferred those features. More direct interpretation would hold an image structure approximately fixed while changing one characteristic—contrast, boundary sharpness, orientation, or spatial period—and measure the sensitivity of the fitted utility. Such counterfactual perturbations remain an important test before interpreting latent directions as scientific concepts.

Figure~\ref{fig:5} summarizes the preference-trained model. For a representative heuristic trace (HDH 1), selected candidates occupy one end of the learned two-dimensional embedding (Fig.~\ref{fig:5}a), the fitted utility increases toward that end (Fig.~\ref{fig:5}b), and the first latent coordinate correlates positively with mean intensity and the intensity percentiles while the second carries the opposite sign (Fig.~\ref{fig:5}c). Applied separately to the six recorded histories, the same architecture yields distinct spatial utility fields on the common image (Fig.~\ref{fig:5}d, e). The resulting maps vary across the GPT-generated, heuristic-driven, and human-participant histories, reflecting differences in their recorded selection preferences. Fitted models and color scales differ between columns, so the gallery compares spatial structure, not reward magnitudes.

Together, these experiments show how to reconstruct and interrogate a stationary effective preference using either predefined morphology descriptors or a learned image representation. The principal observation is heterogeneity: simple features describe several choice processes effectively, while more expressive functions provide alternative descriptions without consistent predictive gains. Poor agreement with a recorded history may have several causes: important image features may be missing, choices may be inconsistent, relevant context may be unrecorded, or the decision-maker’s priorities may have changed. The temporal analyses examine the last possibility using the same recorded candidate sets.

% Learning changing rewards
\section{Learning changing rewards from the same recorded experiment}\label{sec:4}

Human decision-making is rarely constant over the course of an experiment. A typical session by a human operator progresses through several stages: it starts by establishing the baseline performance of the instrument, proceeds to the interpretation of statistically significant features of the sample, moves to unusual features that attract attention out of curiosity, and may then explore features dictated by a specific scientific question. Preferences can also change for other reasons: the operator learns from the information already acquired, alternates between exploring new regions and returning to known ones \cite{ref77,ref78}, or simply becomes tired, and such changes can be gradual or occur abruptly at isolated points \cite{ref79}. Switching between discrete strategies and gradual drift of decision weights are both well documented in behavioral studies \cite{ref80,ref81,ref82}. A single stationary reward fitted to such a history average over several distinct preferences. In this section we model these changes explicitly with two complementary descriptions: a hidden Markov model in which a latent state selects among a small set of recurring rewards (Sections 4.1 and 4.2), and a continuous-state model in which the reward weights evolve gradually from decision to decision (Section~\ref{sec:4.3}). Both are tested on controlled sequences with known changes, following established practice for validating behavioral models \cite{ref83}, and on the original decision histories.

\subsection{Decision Histories May Reflect Multiple Objectives}\label{sec:4.1}

A sequence of microscopy selections may reflect several scientific priorities even though every decision has the same format. An experimentalist might first locate domain walls, then seek well-ordered stripes to establish a reference morphology, and later revisit boundaries to test whether a proposed interpretation generalizes. A stationary reward fitted to the complete history averages across these regimes; the average can predict poorly even if each regime individually follows a simple image-dependent rule. We therefore regard stationarity as a hypothesis to be tested rather than an intrinsic property of a recorded trace. We now ask whether the history is better described by a few recurring preferences or by a reward that changes gradually over time.

The hidden Markov formulation associates each unobserved state \(z_{t}\) with a reward \(R_{k}\)(\(x\)) and hence with a distinct conditional choice distribution over the candidates offered at decision t. The Markov transition matrix specifies how persistent those states are and how likely the model is to return to an earlier state. The states are not labeled in advance; the model learns their reward functions and transition probabilities from the choices. A latent state is an operational hypothesis about selection behavior, not a measurement of attention or psychological intent. It is meaningful only when the states make distinct predictions on comparable candidate sets and when modeling their changes provides a reproducible advantage over a stationary description.

\begin{equation}
\label{eq:14}
P(a_t=i\mid C_t,z_t=k)=\frac{\exp[\beta R_k(x_{t,i})]}{\sum_j\exp[\beta R_k(x_{t,j})]},\quad R_k(x)=\mathbf w_k^T\tilde{\boldsymbol\phi}(x)
\end{equation}

where,

\begin{equation}
\label{eq:15}
P(z_t=k\mid z_{t-1}=j)=A_{jk}
\end{equation}

An HMM requires more validation than a plot of posterior state probabilities. Different coefficient vectors can yield essentially the same candidate rankings, and a flexible model may split a stable strategy into redundant states. We therefore compare state-specific choice distributions, top-ranked candidates, transition persistence, and the held-out likelihood associated with different numbers of states. State labels are exchangeable and must be aligned before calculating recovery accuracy against controlled ground truth. Offline forward–backward smoothing and Viterbi decoding can summarize a complete historical sequence but use later observations; they do not measure advance prediction. In chronological testing, the model must predict each winner with the information available before that choice, and only then update its state probabilities.

Chronological prediction propagates the filtered state distribution \(\pi_{t-1}\) through the transition matrix and mixes the state-specific choice models \cite{ref84,ref85},

\begin{equation}
\label{eq:16}
\begin{aligned}P(a_t\mid C_t,H_{t-1})&=\sum_k\pi_{t\mid t-1}(k)P(a_t\mid C_t,z_t=k),\\ \pi_{t\mid t-1}(k)&=\sum_j\pi_{t-1}(j)A_{jk}\end{aligned}
\end{equation}

after which the observed winner updates the state distribution,

\begin{equation}
\label{eq:17}
\pi_t(k)=\frac{\pi_{t\mid t-1}(k)P(a_t\mid C_t,z_t=k)}{\sum_j\pi_{t\mid t-1}(j)P(a_t\mid C_t,z_t=j)}
\end{equation}

Parameters are estimated by generalized expectation–maximization, alternating forward– backward state responsibilities with weighted fits of Eq.~\eqref{eq:9} \cite{ref86,ref87}. Viterbi decoding then identifies the most likely state sequence over the training decisions \cite{ref88}. Persistence priors, explicit-duration models, and temporally extended options provide natural extensions of this formulation \cite{ref89,ref90,ref91}.

To test state recovery rather than rely on interpretive impressions, we construct controlled sequences from the actual microscopy candidate sets while generating new winner labels from known image-dependent rewards. The teacher coefficients, state identities, and transition times remain hidden from the inference routines and are revealed only for evaluation. We test a fixed reward, a single hidden switch, repeated switching between two familiar rewards, gradual drift, and the appearance of a new strategy after training. These conditions separate three questions: whether a model invents changes when the reward is fixed, whether it recovers recurrent strategies, and whether its state space can adapt when the preference changes in a qualitatively new direction. They are tests of the inverse methodology, not observational evidence of a human changing motivation.

\subsection{Discovering recurring strategies and their switches}\label{sec:4.2}

\begin{figure}[htbp]
  \centering
  \paperfigure{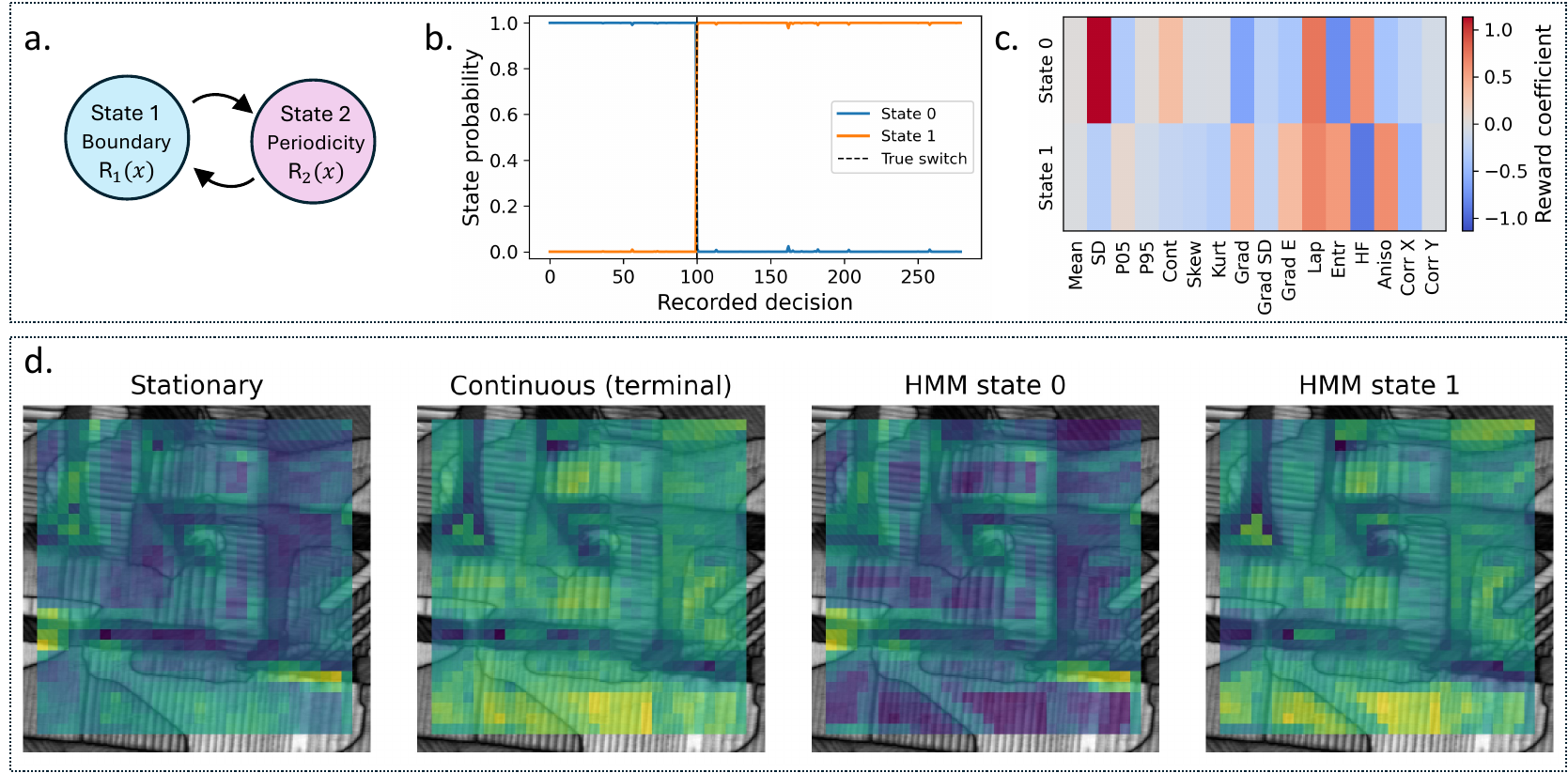}
  \caption{Latent preference states in a constructed two-source decision history. (a) Two-state model: a hidden state selects the reward that governs each recorded choice. (b) Filtered state probabilities over 200 decisions; the dashed line marks the concatenation boundary at decision 100, displayed after fitting. (c) State-specific coefficients of the 16 image descriptors. (d) Spatial utilities of the stationary model, the terminal continuously updated model and the two HMM states.}
  \label{fig:6}
\end{figure}

Figures 6 and 7 examine a constructed 200-decision sequence obtained by concatenating the 100 recorded decisions of GPT 1 and GPT 2. The concatenation introduces a known source boundary after decision 100 without requiring additional microscopy data. A two-state model (Fig.~\ref{fig:6}a) provides a description of the constructed history in terms of two latent reward regimes. The filtered state probabilities are compared with the known concatenation boundary at decision 100, which is displayed only after fitting (Fig.~\ref{fig:6}b). The two states differ mainly in the anisotropy and directional-correlation coefficients, which enter with opposite signs (Fig.~\ref{fig:6}c). Their spatial utilities emphasize different domain regions; the stationary map resembles that of state 0 and the terminal continuous map that of state 1 (Fig.~\ref{fig:6}d). The construction combines distinct generating histories and therefore tests source discrimination rather than a change within one decision-maker.

The more stringent programmed-switch test uses a single generating candidate protocol with two known reward functions and undisclosed transition times. For recurrent changes introduced at decisions 64, 123, 204, and 243, the fitted HMM achieves 61.9\% top-1 accuracy on the chronological held-out segment, compared with 31.0\% for the stationary model and 44.0\% for the continuously updated model. This supports the narrower claim that discrete latent reward regimes can be recovered from sufficiently informative choices under controlled conditions. It does not establish that the original experimentalists changed objectives at those points. The controlled stationary, recurrent-switch, and continuous-drift comparisons are described in Supplementary Section S5.

\begin{figure}[htbp]
  \centering
  \paperfigure{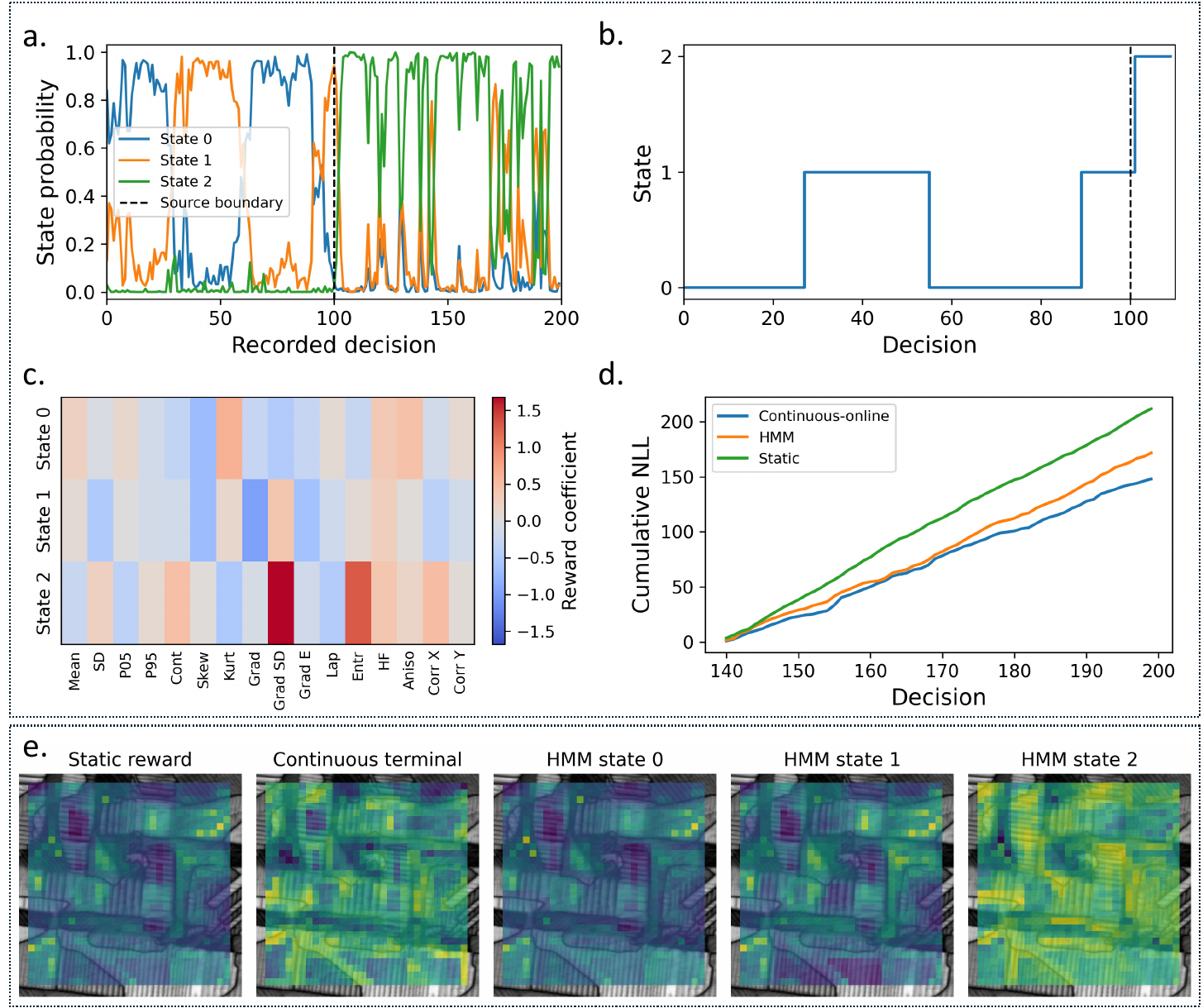}
  \caption{Temporal models on a concatenation of two GPT-generated decision traces. (a) Filtered probabilities of three HMM states; the dashed line marks the known boundary at decision 100, displayed after fitting. (b) Most probable state path over the training prefix (Viterbi). (c) State-specific descriptor coefficients. (d) Cumulative held-out choice NLL of the stationary, HMM and continuously updated models over the chronological test window. (e) Spatial utilities of the stationary model, the terminal continuous model and the three HMM states.}
  \label{fig:7}
\end{figure}

Figure~\ref{fig:7} extends the analysis of the same concatenated GPT 1–GPT 2 sequence by considering a three-state HMM selected using validation likelihood. The filtered probabilities show two alternating regimes before the boundary and a single persistent state after it (Fig.~\ref{fig:7}a), and the training-prefix Viterbi path reproduces this structure (Fig.~\ref{fig:7}b). The states differ most in the kurtosis, mean-gradient and anisotropy coefficients (Fig.~\ref{fig:7}c). Over the chronological test window (decisions 140–199), the HMM and the continuously updated model accumulate a substantially lower NLL than the stationary model (Fig.~\ref{fig:7}d). Because this window lies entirely after the boundary, the advantage reflects adaptation to the second source rather than anticipation of an unseen switch; the programmed-switch experiments address that case. The spatial maps differ only modestly between states (Fig.~\ref{fig:7}e).

The negative controls are as important as the successful switch demonstrations. In exploratory fits to the original traces, two numerical state vectors frequently selected the same top candidate, had small choice-distribution divergence, and failed to outperform a one-state model under approximate state-count criteria. More parameters alone therefore do not establish additional scientific strategies. By contrast, a controlled stationary teacher should yield either one preferred state or negligible evidence for meaningful switching. We also ask whether the inferred strategies reappear with comparable reward rankings after an interruption, because that is the distinctive use case for a Markov representation of recurring objectives.

The previously unseen-strategy condition exposes a second limitation. An HMM fitted only on the first segment of a history cannot immediately represent a reward that never occurred in training. It may assign such decisions to the least-inconsistent existing state or adapt only after collecting further evidence. A continuously evolving reward can sometimes respond more rapidly, but at the cost of representing abrupt recurrent changes poorly or tracking individual noisy choices too closely. These are competing explanatory hypotheses, not a universal competition with an expected single winner. Their relative utility depends on the separation of the generating rewards, transition frequency, number of choices, and information content of the candidate set.

\subsection{Gradual changes as an alternative to discrete strategies}\label{sec:4.3}

Scientific priorities can also change without a discrete transition. As familiarity with a specimen grows, an operator may gradually favor a different spatial frequency, expand acceptable contrast, or shift attention from isolated boundaries to extended patterns. We model this alternative using a reward vector that evolves according to a smooth stochastic process, with each new choice updating the current estimate. The previous estimate predicts the next choice, while the newly observed winner is assimilated only afterward. The process-noise setting controls the trade-off between adaptability and stability. A small value describes nearly stationary preference; a larger value allows rapid changes but is vulnerable to noise. Full filtering equations and the chronological evaluation protocol are given in Supplementary Section S5.

\begin{equation}
\label{eq:18}
\mathbf w_t=\mathbf w_{t-1}+\boldsymbol\epsilon_t,\quad\boldsymbol\epsilon_t\sim\mathcal N(0,q\mathbf I),\quad R_t(x)=\mathbf w_t^T\tilde{\boldsymbol\phi}(x)
\end{equation}

With a Gaussian approximation to the posterior, each observed choice updates the mean and covariance as

\begin{equation}
\label{eq:19}
\begin{aligned}\mathbf w_t&=\mathbf w_{t-1}+\Sigma_t[\tilde{\boldsymbol\phi}(x_{t,a_t})-\bar{\boldsymbol\phi}_t],\\\Sigma_t^{-1}&=(\Sigma_{t-1}+q\mathbf I)^{-1}+\sum_j p_{t,j}(\tilde{\boldsymbol\phi}_{t,j}-\bar{\boldsymbol\phi}_t)(\tilde{\boldsymbol\phi}_{t,j}-\bar{\boldsymbol\phi}_t)^T\end{aligned}
\end{equation}

where \(p_{t,j}\) is the predicted probability of candidate \(j\) and \(\bar\phi_t=\sum_j p_{t,j}\tilde\phi_{t,j}\) is the expected descriptor vector. This is the softmax-observation analogue of a Kalman filter \cite{ref82,ref92}: the update moves the weights toward the descriptors of the observed winner, scaled by the current uncertainty, and q sets the trade-off between stability and adaptability.

\begin{figure}[htbp]
  \centering
  \paperfigure{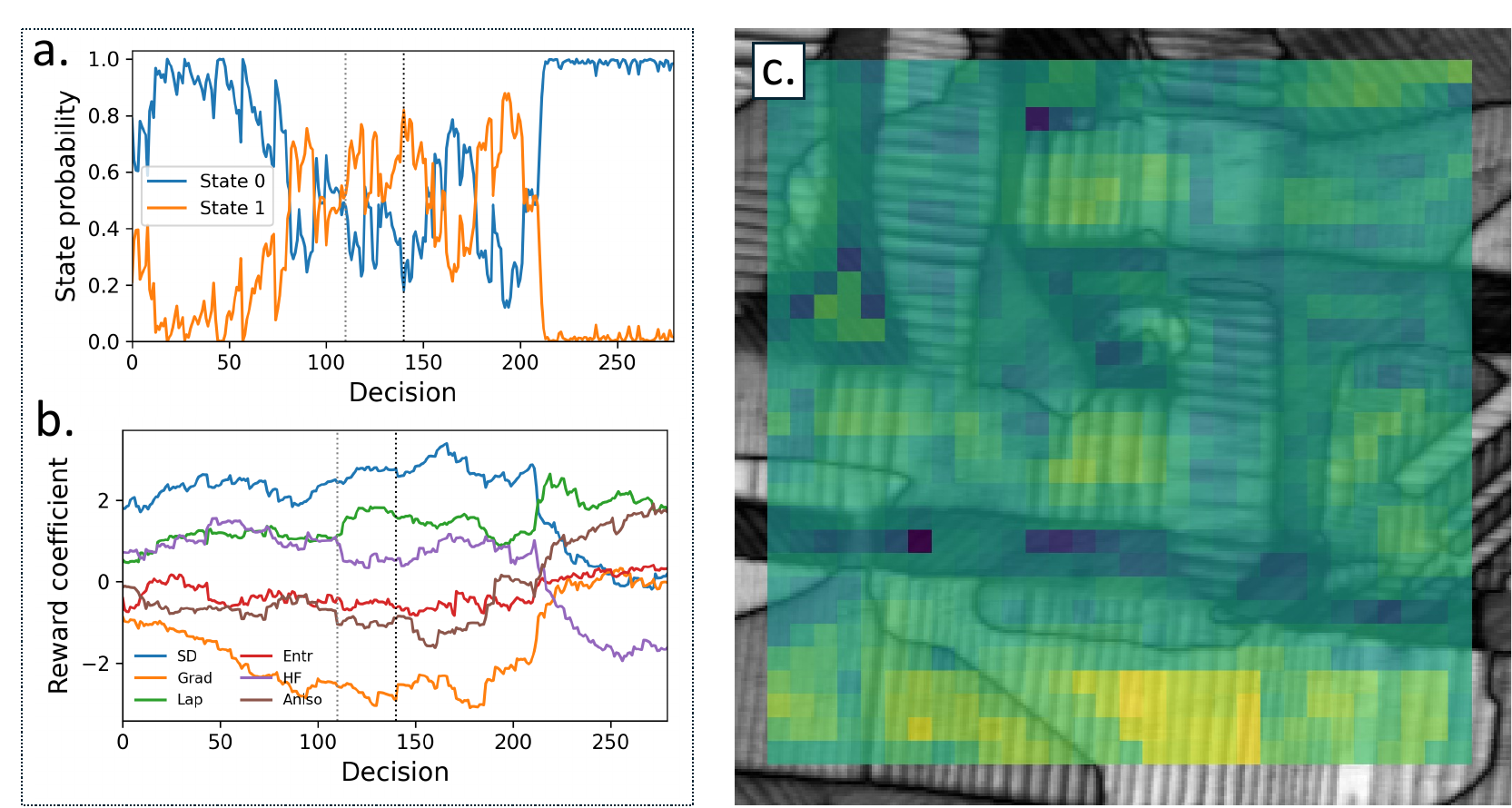}
  \caption{Discrete-state and continuous-state descriptions of a changing preference. (a) Online filtered probabilities of the fitted HMM states after each observed choice in a controlled nonstationary sequence. (b) Continuously updated reward coefficients for the same sequence. Dotted lines in (a) and (b) mark the ends of the training and validation windows. (c) Terminal spatial reward map of a continuously updated model on the archived image.}
  \label{fig:8}
\end{figure}

Figure~\ref{fig:8} contrasts discrete-state and continuous-state descriptions on a controlled nonstationary sequence. The HMM represents changing preferences through probabilities assigned to distinct latent states (Fig.~\ref{fig:8}a), whereas the continuous-state model allows the reward coefficients to evolve over successive decisions (Fig.~\ref{fig:8}b). Dotted lines mark the ends of the training and validation windows (decisions 154 and 196). Fig.~\ref{fig:8}c shows the terminal reward map of a continuously updated model on the archived image, illustrating how a fitted coefficient vector is read out spatially.

The controlled comparison includes stationary choice, one abrupt change, recurring switches, continuous drift, and a new preference introduced after training. Under the saved smooth-drift construction, online continuous updating attains 61.9\% held-out top-1 accuracy, compared with 50.0\% for the HMM and 4.8\% for the stationary model. In the recurrent-switch construction, the HMM achieves 61.9\% top-1 accuracy, compared with 31.0\% for the stationary model and 44.0\% for continuous updating. These differences are evidence about the tested generating processes and chosen hyperparameters, not a theorem establishing one method as intrinsically superior. They demonstrate why the type of nonstationarity is a scientific hypothesis that should be evaluated rather than assumed.

The six unmodified decision histories, comprising two GPT-generated, two heuristic-driven, and two human-participant histories, provide a separate test of whether temporal modeling improves retrospective prediction. In the saved chronological evaluation, mean held-out top-1 accuracy is 40.0\% for the stationary model, 40.0\% for the HMM, and 45.0\% for continuous updating, with corresponding mean NLL values of 2.10, 2.10, and 1.85. The gains are much smaller than in the programmed-switch cases. Several histories are adequately described by one effective state, and some fitted multistate models have little behavioral distinction between their states. These observations do not verify spontaneous changes in a human investigator's motivation. Four of the six recorded histories were generated using GPT-based evaluators or human-defined heuristics, and the study was not designed as a prospective investigation of changes in human motivation.

The same apparent decision change can have several explanations. Candidate populations can shift, image quality can vary, task instructions can change, or the evaluator can acquire information not represented in the recorded patch descriptors. Conditional modeling accounts for the alternatives offered, but it cannot identify omitted contextual variables from image choices alone. Future retrospective datasets should therefore retain instructions, instrument mode, earlier observations, and independently annotated changes of task when these exist. Such records would permit comparison between a changing reward, a changing belief about the specimen under a fixed reward, and a change in admissible actions. The present study establishes methods and controlled recoverability, not a unique psychological reconstruction.

% Discussion
\section{Discussion}\label{sec:5}

\subsection{What the decision histories reveal}\label{sec:5.1}

The analyses show that a recorded decision history is a usable source of experimental objectives. When the alternatives offered at each step are preserved, a reward reconstructed from part of a history can predict withheld choices substantially better than chance. Performance is particularly strong for some heuristic-driven histories, indicating that explicit image-dependent selection rules can be recovered from their recorded decisions. Interpretable descriptors make the inferred preferences readable: the fitted coefficients identify which image characteristics, such as intensity statistics, gradients or anisotropy, distinguish selected patches from their competitors. Learned representations provide complementary descriptions and, in the preference-trained form, relate their latent axes back to the same familiar statistics. Projected over the sample, every model yields a spatial reward field that concentrates value on the domain regions the decision-maker favored. In the temporal setting, the hidden Markov model improved prediction under programmed recurrent switching, while the continuously updated model performed better under the tested gradual-drift condition. Together, these results provide both a framework and a benchmark for learning operator rewards from decision traces.

These results also have clear limits. The inverse problem learns an effective reward defined by observable choice behavior. It is neither a direct measurement of a physical property nor a uniquely recoverable report of scientific intention. Under the conditional softmax model, adding a constant to all candidate utilities leaves the selection probabilities unchanged; reward scale is also confounded with choice temperature when both are free. Highly correlated features can support many nearly equivalent coefficient vectors, and any preference depending on unrecorded context is outside the identifiability of an image-only model. Different rewards may therefore make similar predictions on the observed candidate sets while assigning different scores to patches never offered. The practical question is not whether a single 'true human reward' was recovered, but whether the model supplies a stable, inspectable and transferable account of a particular decision rule. That claim becomes stronger only when it survives held-out choices, altered candidate sets, independent fields of view, and independent evaluation of the resulting scientific observations.

The stationary comparisons show that two choices shape a reconstructed reward: which image properties enter the model and which function turns them into a preference. A linear reward over recognizable descriptors is a transparent starting point and often captures the heuristic strategies well, while nonlinear and learned models offer alternative descriptions that did not, on average, improve prediction in the present benchmark. This reflects the tasks and data studied here rather than evidence that microscopists think linearly. For changing preferences, a shift in the fitted weights is not automatically a change in objective, because correlated descriptors can trade weight without changing which patch is preferred \cite{ref41,ref42,ref93}. The controlled sequences address this by comparing the inferred changes with known ones; establishing changing priorities in real operator histories will require independent records of the operator’s intent, such as instructions or annotations.

\subsection{From a reconstructed reward to what the instrument might do}\label{sec:5.2}

A reward alone does not define how the instrument should act. A reward ranks candidate locations, whereas an acquisition policy also decides which candidates are considered, how much each measurement costs, whether locations can be revisited, and how redundancy is avoided \cite{ref94}. Simply selecting the highest predicted reward over an image tends to return many overlapping patches in one favorable region, so the proposed approach has to be combined with a strategy that maintains diversity, for example by drawing limited candidate sets, adding a coverage-equalization term or using uncertainty estimates. In this paper, all reward maps are computed on the already acquired image; they show where the model would assign value, and their use to drive new measurements remains to be demonstrated. The policy options and their assumptions are outlined in Supplementary Sections S2 and S6.

An independent endpoint is essential when translating inferred preference into autonomous scientific utility. Reusing the model's own reward to grade locations selected by that model would be circular. A stronger test freezes the reward, applies it to new consideration sets or a distinct sample, and evaluates the resulting selections through blinded expert judgments or a physically measurable response. If the purpose is to maximize information gain, the scientific unknown and its observation model must be defined; spatial dispersion alone is not information-theoretic gain. A credible deployment study would match candidate-generation costs and selection budgets while comparing uniform proposals, global greedy selection, best-of-K sampling and explicit reward-plus-coverage policies. Such comparisons remain proposed validation experiments rather than results of the current microscopy archive.

\subsection{Recording decisions changes what can be learned from a laboratory}\label{sec:5.3}

What should a laboratory record to make reward learning possible? Images and spectra alone, even when archived according to FAIR principles \cite{ref95}, are insufficient. A decision trace should preserve the alternatives considered, their identities or coordinates, the selected alternative, time order, instrument conditions, and the instructions or prior information available to the operator. The choice set is particularly important: without it, a selected location cannot be distinguished from a preference under constrained availability. Combined with instrument and sample traces, this information would support reconstruction of recurring operational objectives and detection of changes in what an operator appears to value. It would also permit tests distinguishing a change in reward from a change in physical knowledge or action feasibility. The present experiments show what can be extracted when consideration sets are preserved, while identifying the limitations imposed by omitted context and repeated imaging of one specimen.

Preference learning, expert comparisons, inverse reinforcement learning and human-guided autonomous microscopy provide substantial prior art for inferring utility from choices \cite{ref33,ref34,ref35,ref36,ref37,ref38,ref41,ref42,ref43,ref44,ref45,ref46,ref47,ref48,ref49,ref53,ref96,ref97,ref98,ref99}. Our contribution is not the invention of conditional choice modeling or hidden Markov inference. It is an experimentally motivated formulation in which archived microscopy consideration sets serve as the data for comparing interpretable and learned reward representations, and the same stored decision history can be interrogated for stationary, recurrent, or drifting preferences. The controls separate predictive agreement, recovery of programmed objective changes, and eventual scientific value. The latter remains unverified in this study. Whether such learned objectives generalize across different materials, imaging modalities, operators, and scientific questions is an empirical challenge rather than a consequence of model expressivity.

% Conclusions
\section{Conclusions}\label{sec:6}

The decision history of a microscope operator contains implicit information about human logic and decision-making: which regions are judged worth measuring, relative to which alternatives, and how these judgments evolve during an experiment. This information is rarely written down, yet it is precisely what an automated instrument needs to act on the operator’s behalf.

Recorded microscopy decisions contain recoverable information about effective experimental objectives when the alternatives presented at each choice are preserved. Using a common archived image, we evaluated stationary rewards based on interpretable descriptors, nonlinear functions, and learned image representations. Their predictive value varied substantially among decision histories and added model complexity did not guarantee improved reconstruction. Spatial projections make the fitted utilities inspectable but do not demonstrate that following them produces independent scientific information. Hidden Markov inference recovered concealed recurring reward strategies in controlled image-backed sequences, while continuously updated rewards were more suitable for a tested smooth-drift process. In the unmodified histories, temporal-model improvements were modest, so spontaneous changes of human motivation cannot be inferred. Together, the analyses establish a disciplined route from comparative choices to quantitative experimental preferences: reconstruct a reward, test it on unseen decisions, inspect its physical meaning and identifiability, and evaluate any proposed use through an independent scientific endpoint.

This opens a practical route to automated instrumentation that inherits human expertise. A reward function learned from an operator’s decisions can be frozen and executed by the instrument at machine speed, scoring candidate regions across a sample, prioritizing measurements, and serving as the objective for Bayesian optimization or reinforcement-learning policies. Because the reward is explicit and inspectable, the operator can verify what the instrument values before deployment, correct it, and combine it with physics-based rewards. Monitoring the inferred preference state during operation would further allow the instrument to recognize when the operator’s priorities change and to adapt accordingly. Recording decision histories together with their alternatives should therefore become a routine part of instrument operation, so that each manual session also trains the automated experiments that follow.

% Declarations

\section*{Data and code availability}

The analysis code is organized in a single Jupyter notebook
covering data loading, stationary reward reconstruction,
image-representation learning, and temporal preference
inference. The analysis uses one archived PFM image
(\texttt{scan\_main\_0000.ibw}) and six decision-history
files (\texttt{gpt\_1.pkl}, \texttt{gpt\_2.pkl},
\texttt{HDH\_1.pkl}, \texttt{HDH\_2.pkl},
\texttt{P01.pkl}, and \texttt{P02.pkl}).
The analysis notebook and the data required to reproduce
the reported results are available at
\url{https://github.com/kbarakati/infer-microscopy-objectives}.

\section*{Acknowledgements and competing interests}

This work was supported by the National Science Foundation
through the ATHENA Programmable Cloud Laboratory Test Bed
(Award No. 2607469). All authors contributed to the code
base development via hackathon and post-hackathon analysis
mechanism and contributed to the decision-trace creation.

% Methods
\section*{Methods summary}

The retrospective data contain six 100-decision histories over one 256 × 256-pixel image; each decision records twenty 20 × 20-pixel candidate patches and the winner index. The histories comprise two GPT-generated, two human-defined heuristic and two human-participant traces. A standardized sixteen-descriptor feature vector characterizes local intensity, gradients, Laplacian response, frequency content, entropy, anisotropy, and directional correlation; a six-descriptor subset provides a reduced baseline. Linear, additive-spline, finite random-feature kernel, and iterative models are fitted by regularized conditional-choice optimization and evaluated on complete held-out decision sets. The principal stationary results use a seeded 80/20 random decision split. The preference-trained image model uses a convolutional encoder, a two-dimensional latent space and a finite random-feature utility head, while the distinct earlier VAE analysis learns the image representation before choice modeling. Temporal experiments use chronological training, validation and test windows. In the controlled tests, winner choices are generated from known image-dependent teacher rewards over real candidate sets; teacher parameters and change points are withheld from all fitting procedures. HMM inference uses state-dependent choice emissions and transition probabilities, with state count selected using validation likelihood. Continuous inference updates a regularized preference vector after each newly observed winner. Full protocols, limitations, and code lineage are in Supplementary Information. Models were implemented in Python using scikit-learn, SciPy and PyTorch \cite{ref100,ref101,ref102}; The six-descriptor baseline evaluated here was fitted using the same regularized conditional-choice likelihood as the sixteen-descriptor model. It is distinct from the previously introduced six-descriptor ranking-margin formulation \cite{ref63}.

% ==================================================
% 11. Bibliography
% ==================================================

% Ensure all remaining figures appear before references
\FloatBarrier

% Start references on a new page
\clearpage

\bibliographystyle{unsrt}
\bibliography{references}

% ==================================================
% 12. End document
% ==================================================

\end{document}